\documentclass[twocolumn,prl]{revtex4}%
\usepackage{amsmath}
\usepackage[english]{babel}
\usepackage{graphicx}%
\usepackage{amsfonts}%
\usepackage{amssymb}
\begin{document}

\title{The Damping of the Bose-Condensate Oscillations in a Trap at Zero Temperature}
\author{Yu. Kagan and L. A. Maksimov}
\affiliation{Russian Research Centre Kurchatov Institute, Kurchatov Sq. 1, Moscow, 123182 Russia}

\begin{abstract}
We discuss an existence of the damping for the radial condensate oscillations
in a cylindric trap at zero temperature. The damping is a result of the
parametric resonance leading to energy transfer from the coherent condensate
oscillations to the longitudinal sound waves within a finite frequency
interval. The parametric resonance is due to the oscillations of the sound
velocity. The triggering amplitudes at zero temperature are associated with
the zero-point oscillations

\end{abstract}
\maketitle
\affiliation{

}

\section{Introduction}

The damping of the oscillations of a trapped Bose condensate isolated from
environment is one of the most interesting problems in the physics of the
Bose-Einstein condensation. So far, experimental [1-4] and theoretical [5-7]
investigations have been reduced to studying the damping due to the
interaction of oscillations with thermal normal excitations. The ensemble of
such excitations in essence plays a role of the heat bath. In all cases the
relatively high temperatures $T\gg\hbar\omega_{0}$ have been considered,
$\omega_{0}$ being the frequency of a parabolic trap. However, the principal
question about the origin of the irreversible damping in the oscillating
isolated system at $T=0$ remains open.

In the present work it is shown that such damping at zero temperature, at
least at some definite conditions, really exists. We consider the radial
oscillations of the condensate in an elongated trap with the cylindric
symmetry, which are induced by the rapid decrease of the frequency $\omega(t)$
of the transverse isotropic potential from value $\omega_{0}$ to $\omega_{1}$.
Such statement of the problem has a number of advantages. As is found in [8],
there exists an exact scaling solution of the nonlinear Schr\"{o}dinger
equation for a field operator and its quasiclassical analog, Gross-Pitaevskii
equation, for an arbitrary behavior $\omega(t)$ in the case of a gas in the
isotropic two-dimensional parabolic potential. It is essential that this
solution holds for the quasi-2D case of an elongated trap with the cylindric
symmetry. In particular, a ratio between the interaction energy of particles
and their kinetic energy remains constant. Thus, if the Thomas-Fermi
approximation is applicable in the initial static trap, it remains valid at
all stages of the gas evolution.

The scaling solution describes the space-time evolution of the condensate,
relying only on the solution in the initial static potential of frequency
$\omega_{0}$. After the parameter of the potential takes the value $\omega
_{1}$, the condensate sets into the stable state accomplishing with the radial
oscillations of frequency $2\omega_{1}$. These oscillations are accompanied by
the vibrations of the condensate density and, therefore, by the vibrations of
the sound velocity $c$.

As will be shown below, for a gas in which the sound velocity oscillates, the
phenomenon of parametric resonance appears. The essence of the phenomenon is
that the resonance between the radial condensate vibrations and the
longitudinal wave propagating in the direction of the axial $z$ axis produces
an exponential increment for the amplitude of the wave. At zero temperature
the triggering amplitudes are formed, in essence, by zero-point oscillations.
In the Thomas-Fermi regime, when the chemical potential $\mu\gg\hbar\omega
_{0}$, the sound waves of wavelength $\lambda\approx c/\omega_{1}$ amplifies.
As a result, the dynamic energy of the coherently oscillating condensate
reduces, converting into the energy of the longitudinal sound vibrations. For
the analysis of the initial stage of generation at $T=0$, it is of interest to
find the quantum-mechanical description of the process. Accordingly, in the
last section the rate is found for the elementary process of creating a pair
of longitudinal phonons in a periodic field of the radially oscillating Bose-condensate.

\section{The Bose-condensate oscillations in a trap}

Let us consider a Bose-gas in the elongated cylindric trap with the
symmetrical parabolic potential and longitudinal size $L\gg R_{0}$. Neglecting
edge effects, we can represent the general equation for the Heisenberg field
operator $\hat{\Psi}\left(  \mathbf{r},z,t\right)  $ in the cylindric
coordinate system as
\begin{multline}
i\hbar\frac{\partial\hat{\Psi}}{\partial t}=[-\frac{\hbar^{2}}{2m}\nabla
_{r}^{2}-\frac{\hbar^{2}}{2m}\frac{\partial^{2}}{\partial z^{2}}+\frac{1}%
{2}m\omega^{2}\left(  t\right)  r^{2}]\hat{\Psi}\label{0}\\
+U_{0}\hat{\Psi}^{+}\hat{\Psi}\hat{\Psi}.
\end{multline}
Here $U_{0}=4\pi a\hbar^{2}/m$ where $a$ is the scattering length. The single
simplification in (\ref{0}) is an assumption on the local character of the
interparticle interaction.

A unique property of the above equation, containing parabolic 2D potential of
frequency $\omega(t)$ with an arbitrary dependence on the time, is an
existence of the scaling transformation. This transformation reduces equation
(\ref{0}) to the form in which the frequency of the parabolic potential is
constant and equal to the initial one $\omega_{0}$.

Following [8], let us introduce new scale $\mathbf{\rho}=\mathbf{r}/b\left(
t\right)  $ in the transverse direction and simultaneously define new time
variable $\tau(t)$. Representing the field operator as
\begin{equation}
\hat{\Psi}\left(  \mathbf{r},z,t\right)  =\frac{1}{b\left(  t\right)  }%
\hat{\chi}\left(  \mathbf{\rho},z,\tau\right)  \exp\left[  i\Phi\left(
r,t\right)  \right]  \label{1}%
\end{equation}
we insert this expression into (\ref{0}). In terms of new variables the
equation for operator $\hat{\chi}$ can be reduced to (see Appendix)
\begin{multline}
i\hbar\frac{\partial\hat{\chi}}{\partial\tau}=[-\frac{\hbar^{2}}{2m}%
\nabla_{\rho}^{2}+\frac{1}{2}m\omega_{0}^{2}\rho^{2}-\mu]\hat{\chi}\label{6}\\
+U_{0}\hat{\chi}^{+}\hat{\chi}\hat{\chi}-b^{2}\left(  t\right)  \frac
{\hbar^{2}}{2m}\frac{\partial^{2}\hat{\chi}}{\partial z^{2}},
\end{multline}
if the phase $\Phi$ is expressed via the parameters of transformation
$b\left(  t\right)  $ and $\tau(t)$ by the relation
\begin{equation}
\Phi\left(  r,t\right)  =\frac{mr^{2}}{2\hbar b}\frac{\partial b}{\partial
t}-\frac{\mu\tau(t)}{\hbar},\quad\mu=const, \label{3}%
\end{equation}
and these parameters satisfy the equations
\begin{equation}
\frac{d^{2}b}{dt^{2}}+\omega^{2}\left(  t\right)  b=\omega_{0}^{2}b^{-3}
\label{4}%
\end{equation}%
\begin{equation}
b^{2}\frac{d\tau}{dt}=1. \label{5}%
\end{equation}
Here $\omega_{0}=\omega(-\infty)$ and Eq. (\ref{4}) has the following initial
conditions $b\left(  t\rightarrow-\infty\right)  =1$. In the case of the
problem uniform in the $z$ direction the equation in variables $\rho$, $\tau$
reduces practically to the equations for the static 2D parabolic potential of
frequency $\omega_{0}$. Assuming rapid switching at the initial time moment
for the trap frequency $\omega(t)$ from $\omega_{0}$ to $\omega_{1}$, we find
for the solution of Eq. (\ref{4})
\begin{equation}
b^{2}\left(  t\right)  =\frac{1}{2}\left(  \beta^{2}+1\right)  (1-g\cos
2\omega_{1}t),\;t>0 \label{1511}%
\end{equation}
where $\beta=\omega_{0}/\omega_{1}>1$ and $g=\left(  \beta^{2}-1\right)
/\left(  \beta^{2}+1\right)  $. Thus the parameter $b(t)$ oscillates at
frequency $2\omega_{1}$ within the interval from $1$ to $\beta$.

Substituting this solution into (\ref{5}) and integrating, we obtain
\begin{multline}
\omega_{0}\tau\left(  t\right)  =\omega_{1}t+\arctan[\beta\tan\left(
\omega_{1}t\right)  ]\label{16}\\
-\arctan[\tan\left(  \omega_{1}t\right)  ],\;t>0
\end{multline}
At the stages of small expansion the time $\tau$ varies at the rate closed to
the laboratory one $t$. For the stages of the maximum expansion, the time
$\tau$ in the frame moving together with the gas varies slower.

In the general case the nonlinear operator Schr\"{o}dinger equation (\ref{6})
describes the non-ideal Bose-gas in the concomitant system of coordinates
($\rho$, $\tau$) in which the gas is in the static potential. A single
explicit manifestation of the radial condensate oscillations contains in the
last term on the r.h.s. of (\ref{6}) describing free motion of the gas along
the cylindric axis.

Considering the ground state at $T=0$, we can replace as usually the operator
$\hat{\chi}$ in (\ref{6}) with the macroscopic wavefunction of condensate
$\chi_{0}$. Taking into account that the symmetry of the problem determines an
independence of $\chi_{0}$ on $z$, we arrive at the Gross-Pitaevskii equation,
e.g., [10]
\begin{equation}
i\hbar\frac{\partial\chi_{0}}{\partial\tau}=[-\frac{\hbar^{2}}{2m}\nabla
_{\rho}^{2}+\frac{1}{2}m\omega_{0}^{2}\rho^{2}-\mu]\chi_{0}+U_{0}\chi
_{0}^{\ast}\chi_{0}\chi_{0}. \label{07}%
\end{equation}

From the general representation (\ref{1}) with (\ref{3}) one can conclude that
$\mu$ is a chemical potential corresponding to the static trap of frequency
$\omega_{0}$. In addition, the left-hand side of Eq. (\ref{07}) vanishes and
$\chi_{0}$ in variables $\rho$ and $\tau$ is a stationary real solution of the
equation
\begin{equation}
\lbrack-\frac{\hbar^{2}}{2m}\nabla_{\rho}^{2}+\frac{1}{2}m\omega_{0}^{2}%
\rho^{2}-\mu]\chi_{0}+U_{0}\chi_{0}^{3}=0 \label{11}%
\end{equation}
Let us restrict ourselves by the case when the inequality holds for
\begin{equation}
\mu\gg\hbar\omega_{0} \label{10}%
\end{equation}
and therefore the Thomas-Fermi approximation realizes. Then one can neglect
the ``kinetic energy'', namely, the first term in Eq. (\ref{11}), and obtain
the known solution
\begin{equation}
\chi_{0}=\left(  \frac{\mu}{U_{0}}\right)  ^{1/2}\left(  1-\frac{\rho^{2}%
}{R_{0}^{2}}\right)  ^{1/2}, \label{000}%
\end{equation}
where
\begin{equation}
R_{0}=\sqrt{2\mu/m\omega_{0}^{2}}%
\end{equation}
The number of particles and the energy corresponding to this wavefunction are
equal to
\begin{equation}
N_{0}=\frac{L}{4a}\left(  \frac{\mu}{\hbar\omega_{0}}\right)  ^{2},\quad
E_{0}=\frac{2}{3}\mu N_{0}. \label{12}%
\end{equation}
The energy of the ground state in the trap of frequency $\omega_{1}$ equals
\begin{equation}
E_{1}=\frac{2}{3}\mu_{1}N_{0} \label{13}%
\end{equation}
where
\begin{equation}
\mu_{1}=\mu\frac{\omega_{1}}{\omega_{0}}. \label{14}%
\end{equation}
It is worthwhile that solution (\ref{000}) is valid at the large negative
times when $\omega(t)=\omega_{0}$ as well as at the large positive times when
$\omega(t)=\omega_{1}$. This means that, on one hand, the function $\chi_{0}$
is the wavefunction of the ground state of a gas in the trap of frequency
$\omega_{0}$ and, on the other hand, the same function describes the
condensate state at the large positive times. The state $\chi_{0}$ is static
only in the variables $\rho$ and $\tau$ but in the laboratory frame it
oscillates with the frequency $2\omega_{1}$ as follows from (\ref{1511}). To
describe the space-time evolution of the condensate wavefunction in the
variables $r$ and $\tau$, it is sufficient to substitute (\ref{000}) into the
quasiclassical analog of expression (\ref{1}) and use (\ref{3}), (\ref{1511}),
and (\ref{16}).

\section{Excitations of the Bose-condensate}

With respect to the trap with the parabolic potential of the finite frequency
$\omega_{1}$, the oscillating state $\chi_{0}$ is an excited state. Naturally,
the question arises whether this state is decaying and if it does, what
mechanism is. Let us consider the excited states of the system at the
background of the coherently oscillating condensate. These oscillations can be
found as oscillations of the classical field of the condensate, see, e.g.,
[11]. For this purpose, let us introduce function $\chi=\chi_{0}+\chi^{\prime
}$ instead of operator $\hat{\chi}$ in Eq. (\ref{6}) and linearize it in
$\chi^{\prime}$
\begin{multline}
i\hbar\frac{\partial\chi^{\prime}}{\partial\tau}=[-\frac{\hbar^{2}}{2m}%
\nabla_{\rho}^{2}+\frac{1}{2}m\omega_{0}^{2}\rho^{2}+G-\mu]\chi^{\prime
}\label{17}\\
+G(\chi^{\prime}+\chi^{\prime\ast})-\frac{\hbar^{2}}{2m}b^{2}\left(
\tau\right)  \frac{\partial^{2}\chi^{\prime}}{\partial z^{2}}.
\end{multline}
Here
\begin{equation}
G=U_{0}\chi_{0}^{2}. \label{18}%
\end{equation}
We are interested in the long wavelength longitudinal excitations. For
$k_{z}\equiv k\rightarrow0$, as is clear from the physical reasons, the lowest
branch of the longitudinal excitations is associated with the uniform shift
along the $z$ axis with the radial distribution of the density determined by
the function $\chi_{0}$. Hence this branch is gapless.

For the small finite $k$, the transverse distribution of the density in such
wave changes weakly. This is clearly seen, e.g, from the results obtained in
[12,13] for a static elongated trap. Taking this into account and employing
Eq. (\ref{11}) and definition (\ref{18}), it is easy to find that the first
term on the r.h.s. of (\ref{17}) is obviously small compared with the term
$G\chi^{\prime}$. Then Eq. (\ref{17}) simplifies significantly
\begin{equation}
i\hbar\frac{\partial\chi^{\prime}}{\partial\tau}=G(\chi^{\prime}+\chi^{\prime
}{}^{\ast})-\frac{\hbar^{2}}{2m}b^{2}\left(  \tau\right)  \frac{\partial
^{2}\chi^{\prime}}{\partial z^{2}}, \label{19}%
\end{equation}
We seek for the solution of the equation for the longitudinal excitation in
the form, see, e.g., [11]
\begin{equation}
\chi^{\prime}(\mathbf{\rho},z,\tau)=u(\mathbf{\rho},\tau)e^{ikz}-v^{\ast
}(\mathbf{\rho},\tau)e^{-ikz}. \label{20}%
\end{equation}
Let us introduce the notations
\[
f=\frac{1}{\chi_{0}}\left(  u+v\right)  ,\quad F=\frac{1}{\chi_{0}}\left(
u-v\right)
\]
Inserting (\ref{20}) into Eq. (\ref{19}) for $\chi^{\prime}$ and $\chi
^{\prime}{}^{\ast}$, we find
\begin{align*}
i\hbar\frac{\partial F}{\partial\tau}  &  =\frac{b^{2}\hbar^{2}k^{2}}{2m}f,\\
i\hbar\frac{\partial f}{\partial\tau}  &  =\left(  2G+\frac{b^{2}\hbar
^{2}k^{2}}{2m}\right)  F.
\end{align*}

Let us go over in these equations from variable $\tau$ to the laboratory time
$t$, using relation (\ref{5}). Then the equations can be represented as a
second-order equation
\begin{equation}
\frac{\partial^{2}F}{\partial t^{2}}+\frac{k^{2}}{2m}\left(  \frac{2G}%
{b^{2}\left(  t\right)  }+\frac{\hbar^{2}k^{2}}{2m}\right)  F=0, \label{311}%
\end{equation}
Let us average Eq. (\ref{311}) over $\rho$ and introduce notation
$\left\langle F\right\rangle =\bar{F}$. The average of a product $GF$,
involving that $G$ is a positively determined function, can be represented as
\[
\left\langle GF\right\rangle =\bar{G}\bar{F}%
\]
\newline As a result, we obtain \newline
\begin{align}
\frac{\partial^{2}\bar{F}}{\partial t^{2}}+\Omega_{k}^{2}\left(  t\right)
\bar{F}  &  =0,\label{25}\\
\Omega_{k}^{2}\left(  t\right)   &  =\frac{k^{2}}{2m}\left(  \frac{2\bar{G}%
}{b^{2}\left(  t\right)  }+\frac{\hbar^{2}k^{2}}{2m}\right)  . \label{26}%
\end{align}
\newline For the case of the static potential $b=1$, $\Omega_{k}$ is nothing
else than the frequency of the Bogoliubov spectrum for the longitudinal
excitations, determined for some average condensate density. The sound
velocity equals
\begin{equation}
\bar{c}_{0}=\sqrt{\frac{\bar{G}}{m}} \label{27}%
\end{equation}
\newline 

In the problem which we consider, the scaling parameter $b$ oscillates in
time. As follows from (\ref{26}), this results in the oscillation of the sound
velocity. As will be seen below, we are interested in the longitudinal phonons
of frequencies $\Omega_{k}\simeq\omega_{1}$. Owing to inequality (\ref{10})
this is the purely acoustic region. Accordingly,
\begin{equation}
\Omega_{k}\left(  t\right)  =c\left(  t\right)  k,\;c\left(  t\right)
=\frac{\bar{c}}{b\left(  t\right)  }. \label{28}%
\end{equation}
\qquad Using explicit form of the dependence $b(t)$ (\ref{1511}), we have for
the coefficient $\Omega_{k}^{2}(t)$ in Eq. (\ref{26}) \newline
\begin{equation}
\Omega_{k}^{2}\left(  t\right)  =\frac{\omega_{k}^{2}}{\left(  1-g\cos
2\omega_{1}t\right)  } \label{3144}%
\end{equation}
where
\begin{equation}
\omega_{k}^{2}=\frac{2\left(  \bar{c}k\right)  ^{2}}{\beta^{2}+1} \label{333}%
\end{equation}
Thus, Eq. (\ref{25}) is reduced to the known Hill equation. For the relatively
weak variation of the trap frequency when $\ g\simeq\beta-1<<1$, we arrive at
the Mathieu equation%
\begin{equation}
\frac{\partial^{2}F}{\partial t^{2}}+\omega_{k}^{2}\left(  1+g\cos2\omega
_{1}t\right)  F=0. \label{30}%
\end{equation}
This equation, as well as a more general Hill equation, determines the
parametric resonance connecting the coherent transverse condensate
oscillations at frequency $2\omega_{1}$ with longitudinal phonons with the
frequencies close to $\omega_{1}$. From the physical point of view the
parametric resonance arises due to periodic variation of the condensate
density and, therefore, of the sound velocity (\ref{28}).

\section{The parametric resonance}

To clarify the conditions for appearing the parametric resonance, we restrict
ourselves by the case $g\ll1$ and employ the standard algorithm for solving
Mathieu equation, see e. g. [9]. We seek for a solution as
\begin{equation}
F=A\left(  t\right)  \cos\omega_{1}t+B\left(  t\right)  \sin\omega_{1}t.
\label{030}%
\end{equation}
The coefficients $A(t)$ and $B(t)$ are assumed to be the slowly varying
functions of the time $t$ and on the substitution of (\ref{030}) into the
Mathieu equation (\ref{30}) one should retain only the terms zero and linear
in $g,$ neglecting second derivatives of $A(t)$ and $B(t)$. As a result, we
arrive at a set of two algebraic linear equations with the constant
coefficients. One of the solutions grows exponentially.
\begin{gather}
A\left(  t\right)  ,B\left(  t\right)  \sim e^{\gamma_{k}t},\quad\gamma
_{k}=\frac{1}{2}\sqrt{\left(  \frac{1}{2}g\omega_{1}\right)  ^{2}-s_{k}^{2}%
},\label{352}\\
s_{k}=2(\omega_{k}-\omega_{1}),\nonumber
\end{gather}
provided that the frequency lies within the narrow band near $\omega_{1}$ with
the width
\begin{equation}
\delta\omega=t_{1}^{-1}=\frac{1}{2}g\omega_{1}. \label{351}%
\end{equation}
Within this band a growth of the amplitude of the sound waves takes place and
therefore the energy converts into the longitudinal modes. As the result the
energy of the coherent transverse condensate oscillations decreases. It is
necessary to have at least one mode of the longitudinal sound excitations
within the energy band of the parametric resonance (\ref{351}). The spacing
between the neighbor modes equals $\Delta\omega=2\pi\bar{c}/L$. Since
$R_{0}\simeq2\bar{c}/\omega_{1},$ this condition holds for if there is an
inequality
\begin{equation}
L>2\pi R_{0}/g. \label{3511}%
\end{equation}

Within the narrow band around $\omega_{1}$ a square of the wavefunction
amplitudes (\ref{352}) increases as $\exp2\gamma_{k}t$. At \textbf{\ }%
$T=0$\textbf{\ }the initial energy equals \ the energy of a mode of zero-point
oscillations. Thus the total energy transferred into the longitudinal modes
due to parametric resonance equals
\begin{equation}
E(t)\simeq\sum\limits_{k}\frac{1}{2}\hbar\omega_{k}\left(  e^{2\gamma_{k}%
t}-1\right)  \theta\left(  \delta\omega-\left|  s_{k}\right|  \right)  .
\label{901}%
\end{equation}
\ \ Let us suppose that a sufficient number of modes lies within the band
(\ref{351}) and a sum in (\ref{901}) can be replaced with an integral
\begin{gather}
E(t)=\frac{gL\hbar\omega_{1}^{2}}{16\pi\bar{c}}J(t),\label{902}\\
J(t)=\int\limits_{-1}^{1}dx\left(  \exp\left[  \frac{t}{t_{1}}\sqrt{1-x^{2}%
}\right]  -1\right) \nonumber
\end{gather}

Let us compare the energy (\ref{902}) \ with the energy of the coherent
condensate oscillations. For an instantaneous transition of the trap frequency
from $\omega_{0}$ to $\omega_{1}$, the wavefunction $\chi_{0}$ does not change
its configuration. Hence the energy of the condensate falls from the initial
magnitude $E_{0}$ (see (\ref{12})) to the magnitude
\[
E^{\prime}=\frac{1}{2}\left(  \frac{\omega_{1}^{2}}{\omega_{0}^{2}}+1\right)
E_{0}.
\]
The energy of the vibrational motion $E_{c}$ which the condensate possesses is
equal to a difference between this energy and the energy of the ground
condensate state (\ref{13}) corresponding to frequency $\omega_{1}$
\begin{equation}
E_{c}=E^{\prime}-E_{1}=\frac{1}{3}\mu N_{0}g^{2} \label{9039}%
\end{equation}
As a result, the damping of the condensate oscillations is characterized by a
ratio
\begin{equation}
\frac{E(t)}{E_{c}}=\frac{3a\omega_{0}}{4\pi g\bar{c}}\left(  \frac{\hbar
\omega_{0}}{\mu}\right)  ^{3}J(t)\simeq\frac{a}{gR_{0}}\left(  \frac
{\hbar\omega_{0}}{\mu}\right)  ^{3}J(t). \label{903}%
\end{equation}
For the small times $J(t)\simeq t/t_{1}$, and the ratio (\ref{903}) reduces
to
\begin{equation}
\frac{E(t)}{E_{c}}\simeq\frac{3}{16}\frac{a\omega_{0}^{2}}{\bar{c}}\left(
\frac{\hbar\omega_{0}}{\mu}\right)  ^{3}t. \label{9031}%
\end{equation}
For the times $t\gg t_{1}$, the damping grows exponentially
\begin{equation}
\frac{E(t)}{E_{c}}\simeq\frac{3}{2\pi}\frac{a}{gR_{0}}\left(  \frac
{\hbar\omega_{0}}{\mu}\right)  ^{3}\left(  \frac{2\pi t_{1}}{t}\right)
^{1/2}\exp\frac{t}{t_{1}}. \label{904}%
\end{equation}
The factor in front of the exponential is much smaller compared with unity for
the realistic values of $g$. Hence, within the logarithmic accuracy the
typical time for decaying the condensate oscillations is equal to
\begin{equation}
t_{\ast}=\frac{2}{g\omega_{1}}\ln\left\{  \frac{gR_{0}}{a}\left(  \frac{\mu
}{\hbar\omega_{0}}\right)  ^{3}\left(  \frac{t_{\ast}}{2\pi t_{1}}\right)
^{1/2}\right\}  . \label{912}%
\end{equation}
This time is large compared with the period of condensate oscillations\ not
only due to the smallness of $g$ but also due to the large factor in the
argument of logarithm. Thus the damping of the condensate oscillations due to
the mechanism considered is a rather slow process though appearing inevitably
at $T=0$ under conditions concerned.

The parametric resonance leads to reducing the number of particles in the
condensate. In principle, this could require to solve a self-consistent
problem. However, from the general considerations it is clear that the escape
of particles from the condensate can be neglected if the vibrational energy
per particle $g^{2}\mu$ (see (\ref{9039}) ) is small compared with the
temperature of Bose-condensation $T_{c}$.

\section{The quantum calculation of the phonon generation at the initial stage}

The classical parametric resonance develops from the moment $t=0$ only in the
case when nonzero amplitude of the resonant modes within the interval
$\delta\omega$ (\ref{351}) exists from the very beginning. At the finite
temperature these are amplitudes of thermal oscillations. At $T=0$ the initial
amplitudes should be connected with the zero-point oscillations of the
longitudinal modes. This is explicitly taken into account in the preceding
section. It is of interest to trace the initial stage of the phonon
generation, remaining within the framework of the purely quantum mechanical
consideration. For this purpose, we should return to the operator description
of excitations in the system.

The general equation for operator $\hat{\chi}^{\prime}$ has the form analogous
to that of Eq. (\ref{17})
\begin{align}
i\hbar\frac{\partial\hat{\chi}^{\prime}}{\partial\tau}  &  =h_{0}\hat{\chi
}^{\prime}+G\left(  \hat{\chi}^{\prime}+\hat{\chi}^{\prime+}\right)
-\frac{\hbar^{2}}{2m}b^{2}\left(  \tau\right)  \frac{\partial^{2}\hat{\chi
}^{\prime}}{\partial z^{2}},\\
h_{0}  &  =-\frac{\hbar^{2}}{2m}\nabla_{\rho}^{2}+\frac{1}{2}m\omega_{0}%
^{2}\rho^{2}+G-\mu
\end{align}

We go over from variable $\tau$ to $t$ (see (\ref{5})) and single out the term
associated with the coherent condensate oscillations in the explicit form
\begin{align}
i\hbar\frac{\partial\hat{\chi}^{\prime}}{\partial t}  &  =h\hat{\chi}^{\prime
}+\left[  h_{0}\hat{\chi}^{\prime}+G\left(  \hat{\chi}^{\prime}+\hat{\chi
}^{\prime+}\right)  \right]  \left(  b^{-2}\left(  \tau\right)  -1\right)
,\label{44}\\
h  &  =h_{0}-\frac{\hbar^{2}}{2m}\frac{\partial^{2}}{\partial z^{2}}\nonumber
\end{align}

Assuming that $\left(  b^{-2}\left(  \tau\right)  -1\right)  <<1$, we treat
the last term on the r.h.s. as a perturbation. From the standard commutation
rule for the field operators $\hat{\Psi}\left(  \mathbf{r},z,t\right)  $ and
representation (\ref{1}) it directly follows the commutation rule for
operators $\hat{\chi}^{\prime}$%
\begin{multline*}
\lbrack\hat{\chi}^{\prime}(\mathbf{\rho},z,t),\hat{\chi}^{\prime
+}(\mathbf{\rho}^{\prime},z^{\prime},t)]\\
=b^{2}\delta\left(  \mathbf{r}-\mathbf{r}^{\prime}\right)  \delta\left(
z-z^{\prime}\right)  =\delta\left(  \mathbf{\rho}-\mathbf{\rho}^{\prime
}\right)  \delta\left(  z-z^{\prime}\right)
\end{multline*}
Using this relation, one can readily write the perturbation hamiltonian
governing the behavior of the last term on the r.h.s. of Eq. (\ref{44})
\begin{multline}
H^{\prime}=\left(  b^{-2}\left(  \tau\right)  -1\right) \label{46}\\
\times\int d^{2}\rho dz\left[  \hat{\chi}^{\prime+}\left(  h_{0}+G\right)
\hat{\chi}^{\prime}+\frac{1}{2}G(\hat{\chi}^{\prime}\hat{\chi}^{\prime}%
+\hat{\chi}^{\prime+}\hat{\chi}^{\prime+})\right]
\end{multline}
Consider at first the excitations in the static case when $b=1$. In the
secondary quantization the operator $\hat{\chi}^{\prime}$ can be written using
the eigenfunctions of hamiltonian $h$ as \newline
\begin{gather}
\hat{\chi}^{\prime}=\sum\limits_{nk}\hat{a}_{nk}\chi_{nk},\;\chi_{nk}%
=\frac{e^{ikz}}{\sqrt{L}}\varphi_{n}(\mathbf{\rho}),\label{47}\\
h_{0}\varphi_{n}(\mathbf{\rho})=E_{n}\varphi_{n}(\mathbf{\rho})\nonumber
\end{gather}
Inserting (\ref{47}) into Eq. (\ref{44}), multiplying both sides of the
equation by $\chi_{nk}^{\ast}$, and integrating over $\mathbf{\rho}$ and $z$,
we find
\begin{equation}
i\hbar\frac{d\hat{a}_{nk}}{dt}=\left(  E_{n}+\frac{\hbar^{2}k^{2}}{2m}\right)
\hat{a}_{nk}+G_{nn}\left(  \hat{a}_{nk}+\hat{a}_{nk}^{+}\right)  \label{48}%
\end{equation}
where
\[
G_{nn}=\int d^{2}\rho\varphi_{n}^{2}(\mathbf{\rho})G(\rho)
\]
Equation (\ref{48}) is approximate since we have omitted the terms nondiagonal
in $n$ and proportional to $G_{nn^{\prime}}$. The involvement of these terms
leads to the quantitative corrections alone. Employing the standard Bogoliubov
transformation
\begin{equation}
\hat{a}_{nk}=u_{nk}\hat{b}_{nk}-v_{nk}\hat{b}_{n-k}^{+}, \label{49}%
\end{equation}
and obvious condition
\[
i\hbar\frac{d\hat{b}_{nk}}{dt}=\varepsilon_{nk}\hat{b}_{nk}%
\]
for independent collective excitations, we find the spectrum of excitations
and the values of coefficients $u_{nk},v_{nk}$. For the reasons mentioned in
the preceding section, we consider only the phonon generation corresponding
only to the lowest branch of excitations $n=0$. For $\varepsilon_{nk}$ in this
case, we find the expression coinciding with (\ref{26}) at $b=1$ if $G_{00}$
is understood as $\bar{G}$. Now let us find the phonon generation in first
approximation in $H^{\prime}$ (\ref{46}). For this purpose, using
representation (\ref{47}), we perform transformation (\ref{49}), considering
only the terms with n=0. Since we consider the case T=0, the corresponding
generation proves to be associated only with the terms having a product
$\hat{b}_{0k}^{+}\hat{b}_{0-k}^{+}$ in the hamiltonian, i.e., with the
production of two phonons with the opposite momenta. Correspondingly we find
directly for the transition amplitudes
\begin{multline}
A_{k,-k}=\frac{1}{2}g\bar{G}\cos\left(  2\omega_{1}t\right)  (u_{0k}%
-v_{0k})^{2}\\
\times\langle k,-k\left|  \hat{b}_{0k}^{+}\hat{b}_{0-k}^{+}\right|  0\rangle
\end{multline}
Thus, at the initial stage of relaxation the probability of generating phonon
pairs per unit time is equal to
\begin{equation}
W_{k,-k}=\frac{\pi}{8\hbar}\left[  g\bar{G}(u_{0k}-v_{0k})^{2}\right]
^{2}\delta(2\hbar\omega_{1}-2\varepsilon_{0k}).
\end{equation}
Using the known relations for the coefficients of the Bogoliubov
transformation, e.g., [10], we have for the region of sound excitations (see
(\ref{333}))%
\[
(u_{0k}-v_{0k})^{2}\simeq\frac{\varepsilon_{0k}}{\mu},\;\varepsilon_{0k}%
=\hbar\omega_{k}%
\]
Hence we readily find the power of the energy lost by the oscillating
condensate
\begin{equation}
\dot{E}=\frac{L}{2\pi\hbar\bar{c}}\int d\varepsilon2\varepsilon W_{k,-k}%
=\frac{L\hbar}{64\bar{c}}g^{2}\omega_{1}^{3}.
\end{equation}
This result coincides with that (\ref{9031}) (taking into account (\ref{9039})
and (\ref{12})) obtained in the preceding section for the parametric resonance
at the initial times $t\ll t_{1}.$ The quantum mechanical calculation confirms
our notion on generating the longitudinal waves at zero temperature as an
enhancement of zero-point oscillations due to parametric resonance.

\section{The parametric resonance at the multiple frequencies}

As is known, the Mathieu equation (\ref{30}) results in the parametric
resonance at the frequencies close to the multiple frequencies $p\omega_{1}$.
However, considering the phonon generation at these frequencies, we must
return to the general equation (\ref{25}) even in the case $g\ll1$. The
analysis of the equation shows that the parametric resonance is absent for the
even multiple frequencies. The phonon generation can occur only near the
multiple frequencies $p\omega_{1}$ with odd $p$. As $p$ grows, the growth
decrement $E(t)$ and the width of the resonant interval decrease strongly at
any value of $g$. This can easily be traced, analyzing the nearest multiple
resonance of $p=3$. Such unusual result is due to specific behavior of the
oscillating coefficient in Eq. (\ref{25}) varying periodically in time.

\section{Conclusive remarks}

Thus, analyzing the coherent radial oscillations of the condensate in a
cylindric trap, we provide the evidence for the existence of damping of these
oscillations at zero temperature. The damping is a consequence of the
parametric resonance caused by the sound velocity oscillations. As a result
the energy transfer from radial condensate oscillations to longitudinal phonon
modes is realized.

At the finite temperature $\hbar\omega_{1}<T<<\mu$ the same resonance
amplification takes place but now the initial number of phonons with the
energy $\hbar\omega_{1}$ equals $T/\hbar\omega_{1}$ and the initial energy of
the resonance modes equals $\hbar\omega_{1}\cdot(T/\hbar\omega_{1})$. For this
case, in Eq. (\ref{901}) the quantity $\hbar\omega_{1}$ should be replaced
with $T$. Such replacement increases insignificantly the argument in logarithm
(\ref{912}).

Let us estimate the characteristic damping time $t_{\ast}$ for quantum gases
$Rb$ and $Na$. We consider an elongated cylindric trap with $\nu_{0}=400Hz$
$(Ru)$ or $\nu_{0}=10^{3}Hz$ $(Na),$ $L=2\cdot10^{5}$ and assume that the
dimensionless parameter $g\approx0.15$. Then from Eq. (\ref{12}) we find
$\mu\approx90nK$ for the $Rb$ gas. Accordingly, the Thomas-Fermi radius equals
$R\approx1.7\cdot10^{-4}cm.$ For the parameter $t_{1}$ (\ref{351}), taking
into account the relation $\omega_{1}=\left(  1-g\right)  \cdot\omega_{0},$ we
have $t_{1}\approx6.2\cdot10^{-3}s$. The direct calculation estimates the
logarithmic factor in Eq. (\ref{912}) as $\sim10$. Thus, evaluating the
damping time, we find $t_{\ast}\leq0.1s$. Since it is smaller than the
ordinary lifetime of the system, this time is realistic for revealing the damping.

The similar results can be obtained for sodium. In this case the estimates
give $\mu\approx140nK,$ $R\approx1.6\cdot10^{-4}cm,$ and $t_{1}\approx
2.5\cdot10^{-3}s.$ As a result, $t_{\ast}\approx0.03s.$ It is important that
in both cases the condition (\ref{3511}) proves to be fulfilled. At the same
time $g\mu\ll T_{c},$ and this means that the depletion of the condensate is
quite small after the complete damping.

This work is supported by the Russian Foundation for the Basic Researches and
by the Netherlands Organization for Scientific Research (NWO).

\end{document}